\documentclass[journal]{IEEEtran}

\usepackage[utf8]{inputenc}
\usepackage{amsmath}
\usepackage{graphicx}
\usepackage{hyperref}
\usepackage{mathptmx}

\renewcommand{\vec}[1]{\mathbf{#1}}
\newcommand{\abs}[1]{\lvert#1\rvert}
\newcommand{\norm}[1]{\lVert#1\rVert}
\DeclareMathOperator{\Real}{\mathrm{Re}}
\DeclareMathOperator{\Imag}{\mathrm{Im}}

\begin{document}
\title{The Effect of Metal Thickness on Si Wire to Plasmonic Slot Waveguide Mode Conversion}
\author{\c{S}\"ukr\"u~Ekin~Kocaba\c{s} \thanks{Department of Electrical and Electronics Engineering at Ko\c{c} University, Istanbul, Turkey. ekocabas@ku.edu.tr}\thanks{Preprint version of the online copy available at \url{http://dx.doi.org/10.19113/sdufbed.88342} for the ``Süleyman Demirel Üniversitesi Fen Bilimleri Enstitüsü Dergisi."}}
\maketitle

\begin{abstract}
We investigate mode converters for Si wire to plasmonic slot waveguides at 1550 nm telecom wavelength. The structures are based on a taper geometry. We provide optimal dimensions with more than 90\% power transmission for a range of metal (Au) thicknesses between 30-250 nm. We provide details on how to differentiate between the total power and the power in the main mode of the plasmonic slot waveguide. Our analysis is based on the orthogonality of modes of the slot waveguide subject to a suitable inner product definition. Our results are relevant for lowering the insertion loss and the bit error rate of plasmonic modulators.
\end{abstract}


\section{Introduction} \label{sec:intr}
The practice of using waveguides to transfer information between different points in space with high bandwidth photonic links is gaining popularity. Fiber optics have replaced electrical wiring for a range of distances from thousands of kilometers coast to coast long-haul connections, to meter scaled rack to rack communications in data centers. The burgeoning field of silicon (Si) photonics has made it possible to build integrated opto-electronic components for a more intimate and efficient coupling of electronics for data processing and optics for data transfer.

Si photonic links are based on the Si wire waveguides which are typically fabricated on silicon on insulator (SOI) wafers \cite{Yamada2011a}. In addition to waveguides, photonic links also require light sources, modulators for electrical to photonic and photodetectors for photonic to electrical conversion of information. Mach-Zehnder interferometers and resonant microring cavities have been used in Si modulators \cite{Thomson2016}. Both technologies have relatively large power consumption. In order to have energy efficient and small footprint modulators, recent designs incorporated metallic (i.e.\ plasmonic) and Si wire waveguides \cite{Melikyan2014,Haffner2016}. The coupling rate from Si wire to plasmonic slot waveguides is an important parameter that contributes to the insertion loss and hence the bit error rate of the modulators \cite{Hoessbacher2017}.

Previous studies on Si wire to plasmonic slot waveguide transitions focused mostly on relatively thick, 180-250 nm Au layers \cite{Tian2009,Han2010,Chen2015,Zhu2016a}. A recent study focused on layers of 20-30 nm thickness \cite{Ono2016a}. The propagation length of the modes in plasmonic slot waveguides changes as a function of the slot dimensions \cite{Veronis2007a}. There is a trade-off between the amount of field enhancement due to the small size of the slots and the propagation distance of the ensuing modes.  It would be advantageous to choose the necessary field enhancement based on the light propagation constraints of the application at hand. Field enhancement and metal layer thickness are two closely coupled variables. In this work, we cover a range of Au layer thicknesses between 30-250 nm and provide blueprints for coupling geometries that work at 1550 nm telecom wavelength range with more than 90\% coupling efficiency. We use numerical simulations to verify our designs and we give details of our modeling techniques which use both simple power extraction as well as modal decomposition of scattered fields for estimating the transmission efficiency of the couplers.

%%%%%%%%%%%%%%%%%%%%%%%%%%%%%%%%%%%%%%%%%%%%%%%%%%%%%%%%%%%%%%%%%%%
\section{Material and Method} \label{sec:method}
\begin{figure}[h]
\begin{center}
	\includegraphics{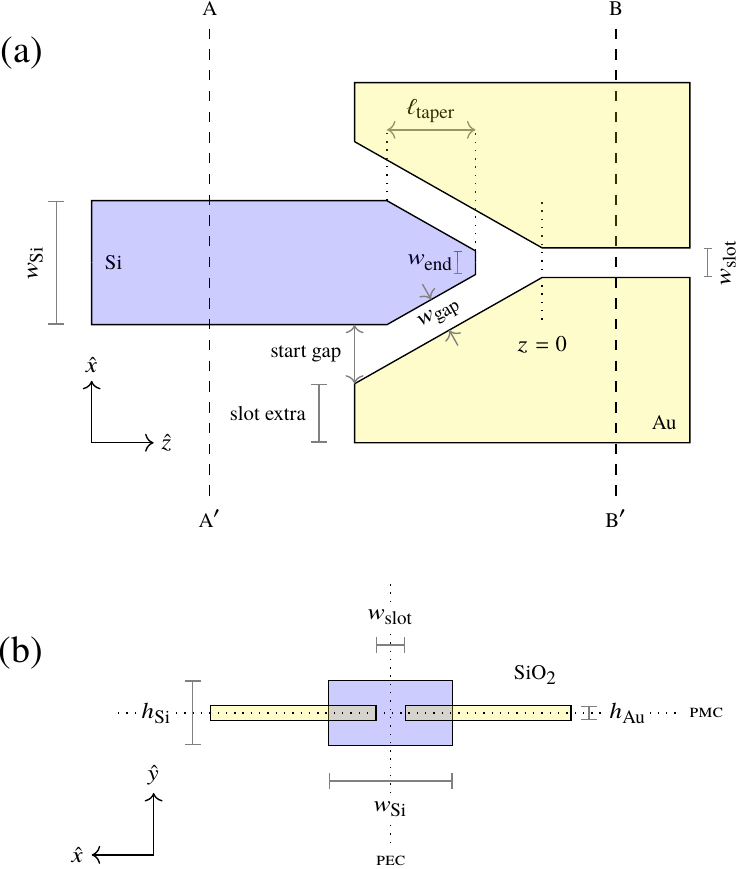}
\end{center}	
	\caption{Geometry and the definitions of the parameters of the mode coupler. a) Top view, b) Cross section view superposing the cuts at the $A-A'$ and $B-B'$ planes.}
	\label{fig_sketch}
\end{figure}

The mode coupler parameters that we employ are illustrated in Fig.\ \ref{fig_sketch}. The aim is to convert the Si wire mode on the left to the mode of the plasmonic slot waveguide on the right, through the taper region at the center. As shown in Fig.\ \ref{fig_sketch}(b), the Si wire and the slot waveguide are centered with respect to one another in the $\hat{y}$ direction. The ambient environment is $\text{SiO}_2$. Such vertical alignment is possible with clean room fabrication techniques as discussed in \cite{Kewes2016}.

We obtain the optical properties of Si from the Sellmeier-type dispersion formula quoted in \cite{Palik1985}. The refractive index of glass is obtained from Eq.\ (20) in \cite{Kitamura2007}. The permittivity of Au is obtained from the supplemental data provided in \cite{McPeak2015}. Since we will be working at a wavelength of $\lambda=1550$ nm, we quote the relevant optical constants of Si, $\text{SiO}_2$ and Au in Table \ref{table:permittivity}. We use the $\exp(+i\omega t)$ convention, thus $\Imag(\epsilon_\text{Au})<0$.

\begin{table}[h]
\begin{center}
\caption{Permittivities of Si, $\text{SiO}_2$ and Au at 1550 nm.}\medskip \label{table:permittivity}
\begin{tabular}{cc}
\hline
 Material& Relative Permittivity $(\epsilon/\epsilon_0)$\\ 
 \hline
  Si & $12.085$ \\
  $\text{SiO}_2$ & $2.0852$ \\
  Au & $-126.80-5.3664 i$ \\ \hline
\end{tabular}
\end{center}
\end{table}

\begin{figure}[h]
	\begin{center}
		\includegraphics{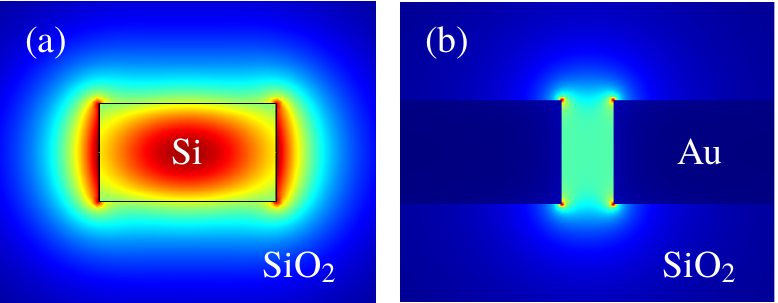}
	\end{center}	
	\caption{Plots of $\norm{\vec{E}}$ for (a) Si wire and (b) plasmonic slot waveguide modes.}
	\label{fig_modes}
\end{figure}

We use the finite element method implementation of the \textsc{comsol} package (v5.1) to solve for the waveguide modes of the Si wire and plasmonic slot waveguides as a function of the height of Si ($h_\text{Si}$) and Au  ($h_\text{Au}$) for fixed values of Si width ($w_\text{Si}$) and slot width ($w_\text{slot}$). Typical mode profiles for $\norm{\vec{E}}$ are plotted in Fig.\ \ref{fig_modes}. The Si wire and plasmonic slot waveguide modes are TE-like with the $E$-field primarily in the $\hat{x}$ direction \cite{Veronis2007a,Yamada2011a}. The plasmonic slot waveguide concentrates the fields within the slot region. The Si wire guides light primarily within the Si region, with evanescent waves leaking to the $\text{SiO}_2$ region.

\begin{figure}[h]
	\begin{center}
		\includegraphics{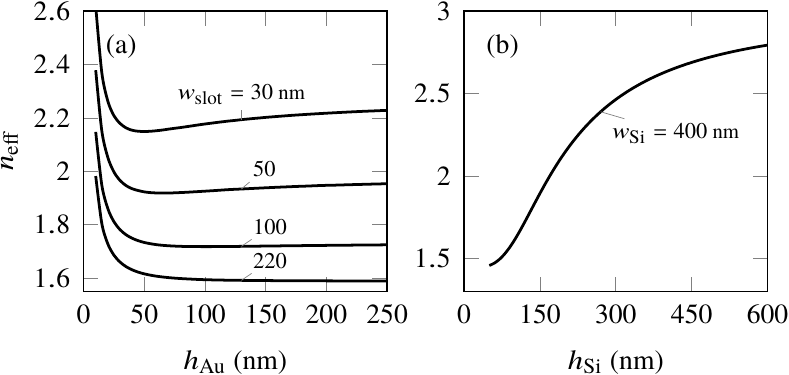}
	\end{center}	
	\caption{Effective index ($n_\text{eff}$) at $\lambda=1550$ nm as a function of waveguide height for (a) plasmonic slot waveguide with $w_\text{slot} = {30,50,100,220}$ nm and (b) Si wire waveguide with $w_\text{Si}=400$ nm.}
	\label{fig_dispersion}
\end{figure}

\begin{figure}[h]
	\begin{center}
		\includegraphics{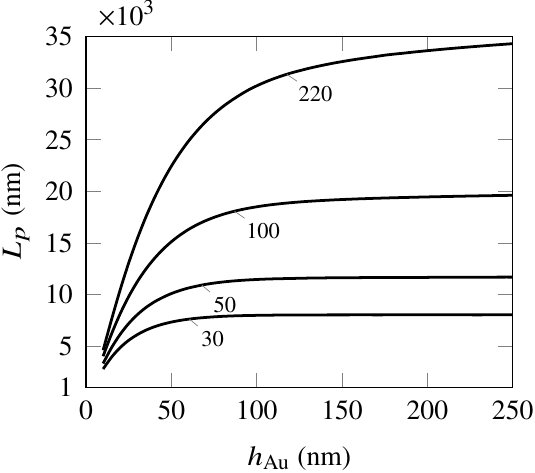}
	\end{center}	
	\caption{Mode intensity propagation length ($L_p$) at $\lambda=1550$ nm as a function of waveguide height for plasmonic slot waveguide with $w_\text{slot} = {30,50,100,220}$ nm.}
	\label{fig_loss}
\end{figure}

The effective index ($n_\text{eff}$) of the modes are plotted in Fig.\ \ref{fig_dispersion} as a function of waveguide height. At large heights, the plasmonic slot waveguide modes approach the 2D metal-insulator-metal modes \cite{Veronis2007a} which have larger $n_\text{eff}$ for smaller slot widths. The price to pay for the large $n_\text{eff}$ is a reduced propagation length ($L_p$) as shown in Fig.\ \ref{fig_loss}. $L_p$ is for the mode intensity and is given by $\frac{-1}{2\Imag(k_z)}$ where $k_z$ is the wave vector of the mode in the direction of propagation. For the Si wire waveguide, the mode approaches the 2D slab waveguide mode as the height increases.

We use \textsc{comsol} to simulate the 3D geometry as depicted in Fig.\ \ref{fig_sketch}. We surround the simulation volume by perfectly matched layers. We source the Si wire waveguide mode from the left and solve for the fields at $\lambda = 1550$ nm, as we vary various geometric variables. We take advantage of the symmetry of the waveguide modes and of the geometry in the $x{-}y$ plane, by putting perfect electric conductor (\textsc{pec}) and perfect magnetic conductor (\textsc{pmc}) boundary conditions at planes that vertically and horizontally bisect the waveguides, respectively [see Fig.\ \ref{fig_sketch}(b)]. These boundary conditions enable us to reduce the number of unknowns by 4-fold, speeding up the simulations.

\begin{figure}[h]
	\begin{center}
		\includegraphics{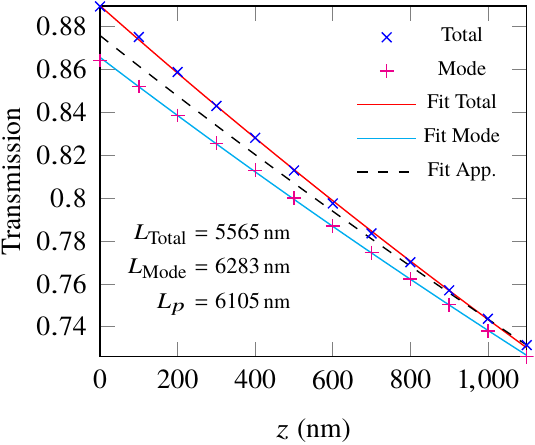}
	\end{center}	
	\caption{Power transmission factor as a function of $z$ for a slot waveguide of $h_\text{Au} = w_\text{slot} = 30$ nm. Different calculation methods are explained in the text.}
	\label{fig_taper}
\end{figure}

We record the tangential $(E_x,E_y,H_x,H_y)$ fields along the straight section of the plasmonic slot waveguide, at different $z$ cuts, starting from $z=0$ as shown in Fig.\ \ref{fig_sketch}(a). We calculate the time averaged total power by numerically integrating the $z$ component of the Poynting vector, $\frac{1}{2}\Real(\vec{E}\times\vec{H}^*)_z =\frac{1}{2}\Real(E_x H_y^*-E_y H_x^*)$ at different $z$ cuts, as shown by the blue `$\times$' symbols in Fig.\ \ref{fig_taper}. Fitting an exponential to these data points (red curve) give us a propagation length $L_\text{Total}=5565$ nm, which is smaller than $L_p=6105$ nm for a slot waveguide of $h_\text{Au} = w_\text{slot} = 30$ nm.

The fields along different $z$ cuts of the slot waveguide include non-bound scattered fields as well. In order to determine the power in the bound mode, we make a modal expansion of the fields. The total electric and magnetic fields at different $z$ cuts, $\vec{E},\vec{H}$ can be expressed as
\begin{align}
\vec{E} = &\alpha \vec{E}_0 + \sum_n \beta_n \vec{E}_n,\label{eq_totE} \\ 
\vec{H} = &\alpha \vec{H}_0 + \sum_n \beta_n \vec{H}_n, \label{eq_totH}
\end{align}  
where $\vec{E}_0,\vec{H}_0$ are the waveguide modes of the slot waveguide as illustrated in Fig.\ \ref{fig_modes}(b) and $\vec{E}_n,\vec{H}_n$ are a decomposition of the scattered fields. The coefficient $\alpha$ is what we are after. We can obtain the power in the bound mode by calculating
\begin{align}
& \frac{1}{2}\Real\left[ \iint (\alpha \vec{E}_0 \times \alpha^*\vec{H}_0^*)_z\ dx dy \right] \nonumber \\ 
& = \abs{\alpha}^2 \frac{1}{2}\Real\left[ \iint (E_{x0}H_{y0}^* - E_{y0}H_{x0}^*)\ dx dy \right].\label{eq_modePow}
\end{align}
In order to obtain the $\alpha$ coefficient, we make use of the orthogonality property of the bound modes that can be proved through the use of the Lorentz theorem. We define an inner product between two sets of fields $(\vec{E}_1,\vec{H}_1)$ and $(\vec{E}_2,\vec{H}_2)$ similar to the one in \cite{Yang2016a} as
\begin{align}
(\vec{E}_1,\vec{H}_1) \cdot (\vec{E}_2,\vec{H}_2) \equiv \frac{1}{2}\iint (\vec{E}_1\times\vec{H}_2 + \vec{E}_2\times\vec{H}_1)_z\ dxdy.
\end{align}
With this definition, we can take the inner product of the fields from a fixed $z$ cut in the 3D simulation \eqref{eq_totE}-\eqref{eq_totH} with the bound modes calculated before, $(\vec{E}_0,\vec{H}_0)$, and use the fact that the bound modes are orthogonal to the non-bound modes, i.e.\ $(\vec{E}_0,\vec{H}_0) \cdot (\vec{E}_n,\vec{H}_n)=0$ to get
\begin{align}
\alpha = \frac{(\vec{E},\vec{H}) \cdot (\vec{E}_0,\vec{H}_0)}{(\vec{E}_0,\vec{H}_0) \cdot (\vec{E}_0,\vec{H}_0)}. \label{eq_alpha}
\end{align}
We calculate $\alpha$ by numerically integrating the relevant fields in \eqref{eq_alpha}. We calculate the power in the bound modes from \eqref{eq_modePow}, and plot them as magenta `$+$' symbols in Fig.\ \ref{fig_taper}. As expected, the power in the bound mode of the slot waveguide is less than the total power in a given $z$-cut. When we fit an exponential to the power in bound modes (cyan curve), we get a decay length of $L_\text{Mode}=6283$ nm, a result much closer to $L_p$. We estimate the power conversion efficiency of a given taper geometry by calculating the power at the $z=0$ cut. We use the total power at $z=1100$ nm, and estimate the power at $z=0$ by using the $L_p$ value from mode analysis, see the black dashed lines in Fig.\ \ref{fig_taper}. This is an approximate method, albeit a practical one, which gives a close estimate of the total power in the bound mode of the waveguide at $z=0$ (slightly less than 88\% in this example).

%%%%%%%%%%%%%%%%%%%%%%%%%%%%%%%%%%%%%%%%%%%%%%%%%%%%%%%%%%%%%%%%%%%
\section{Results} \label{sec:results}
We used the reported values in \cite{Chen2015} and \cite{Ono2016a} for mode conversion to plasmonic slot waveguides with 250 nm and 30 nm thick Au layers, respectively. We searched around these design points, varied $h_\text{Si}$ and other dimensions to come up with the optimized parameters in Table \ref{table:params}.

\begin{table}[h]
\begin{center}
\caption{Optimal dimensions for $h_\text{Au}=30$ \& $250$ nm}\medskip \label{table:params}
\begin{tabular}{ccc}
\hline
 Parameter Name &  Set 1 (nm) & Set 2 (nm) \\ \hline
 $h_\text{Au}$ & 30 & 250 \\
 $h_\text{Si}$ & 300 & 725 \\
 $w_\text{Si}$ & 400 & 400 \\
 $w_\text{slot}$ & 30 & 250 \\
 $\ell_\text{taper}$ & 600 & 1700 \\
 $w_\text{end}$ & 0 & 0 \\
 $w_\text{gap}$ & 20 & 75 \\
 start gap & 200 & 200 \\
 slot extra & 200 & 200 \\
 \hline
\end{tabular}
\end{center}
\end{table}

\begin{figure}[h]
	\begin{center}
		\includegraphics{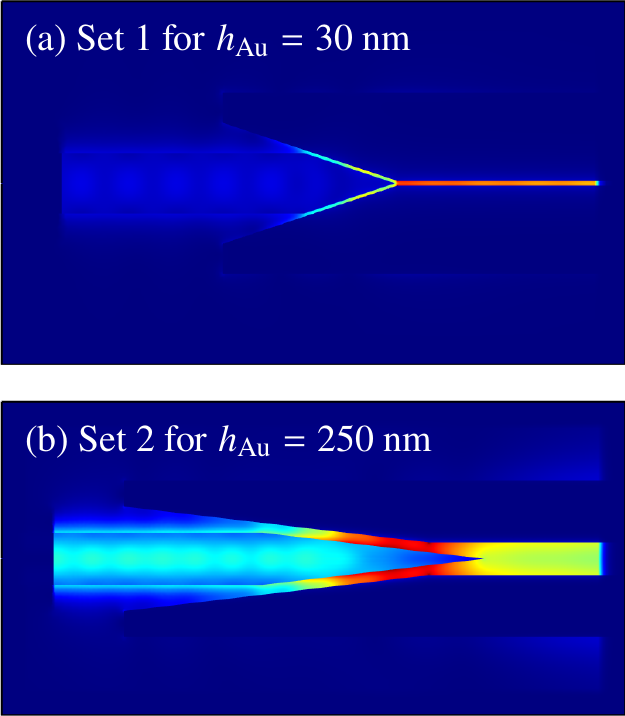}
	\end{center}	
	\caption{Plot of $\norm{\vec{E}}$ through the central plane.}
	\label{fig_fields}
\end{figure}

The norm of the electric field on the central plane that cuts through the structure is plotted in Fig.\ \ref{fig_fields}. The $h_\text{Au} = 30$ nm case has a shorter $\ell_\text{taper}$, and the field intensities in the slot region are higher due to the 30 nm width of the slot. This case has a transmission factor of $\sim$88\%. The $h_\text{Au} = 250$ nm case has a longer taper and a larger slot width of 250 nm with a transmission factor of $\sim$95\%. It is noteworthy that both sets have relatively thick Si layers, $h_\text{Si} \gg h_\text{Au}$. 

\begin{figure}[h]
	\begin{center}
		\includegraphics{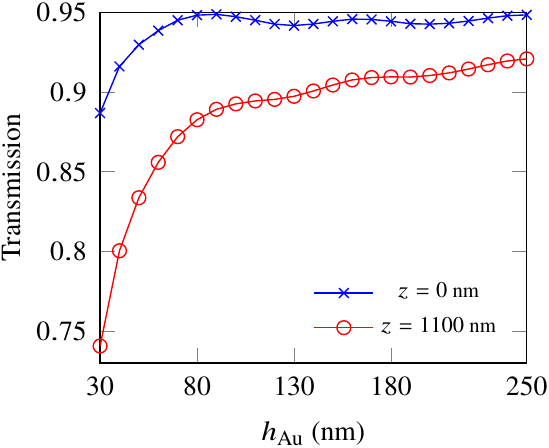}
	\end{center}	
	\caption{Power transmission factor as a function of Au thickness.}
	\label{fig_sweep}
\end{figure}

After we had the optimal values for two different Au thicknesses, we linearly interpolated all the geometry variables in between the two sets listed in Table \ref{table:params} and calculated the transmission factor of the resulting mode converter structures corresponding to Au thicknesses ranging from 30 to 250 nm. The results are provided in Fig.\ \ref{fig_sweep}. We measured the power at the $z=1100$ nm cut, and back propagated to $z=0$ by multiplying the results with $\exp(1100/L_p)$ where $L_p$ values are calculated separately for each $(h_\text{Au},w_\text{slot})$ pair similar to Fig. \ref{fig_loss}.

\section{Discussion and Conclusion} \label{sec:conc}
We investigated optimal structures for coupling the mode of a Si wire waveguide to the mode of a plasmonic slot waveguide. We concentrated on the taper geometry and came up with designs that have over 90\% power transfer efficiency (approaching 95\% in some instances) for Au thicknesses ranging from 30--250 nm. The results that we quote can find applications in compact plasmonic modulator designs with minimal insertion loss and low bit error rates.

Although we focused on Si wire waveguides in this study, the techniques that we present can easily be applied to Si nitride waveguides that have been shown to be highly advantageous for non-linear applications \cite{Ji2017}. Lastly, although our focus has been on taper structures, resonant stub-like elements are another route for designing mode converters as has been demonstrated in 2D structures \cite{Veronis2007}.

%%%%%%%%%%%%%%%%%%%%%%%%%%%%%%%%%%%%%%%%%%%%%%%%%%%%%%%%%%%%%%%%%%%%%%%%
%\bibliographystyle{IEEEtran}
%\bibliography{SiPhotonics}

\begin{thebibliography}{10}

	\bibitem{Yamada2011a}
	Yamada, K. 2011 {Silicon Photonic Wire Waveguides: Fundamentals and
		Applications}. ss 1--29. Lockwood, D. J., Pavesi L., ed. 2011. 
	\href{https://doi.org/10.1007/978-3-642-10506-7_1}{Silicon Photonics II: Components and Integration, Springer, Berlin, 253s.}
	
	
	\bibitem{Thomson2016}
	Thomson, D., Zilkie, A., Bowers, J.~E., Komljenovic, T., Reed, G.~T., Vivien, L., Marris-Morini, D., Cassan, E., Virot, L., Fédéli, J.-M., Hartmann, J.-M., Schmid, J.~H., Xu, D.-X., Boeuf, F., O’Brien, P., Mashanovich, G.~Z.,	Nedeljkovic, M. 2016. Roadmap on silicon photonics. \href{http://stacks.iop.org/2040-8986/18/i=7/a=073003}{{Journal of Optics}, 18(2016), 073003.}
	
	
	\bibitem{Melikyan2014}
	Melikyan, A., Alloatti, L., Muslija, A., Hillerkuss, D., Schindler, P.C., Li, J., Palmer, R., Korn, D., Muehlbrandt, S., Thourhout, D. V., Chen, B., Dinu, R., Sommer, M., Koos, C., Kohl, M., Freude, W., Leuthold J. 2014. High-speed plasmonic phase modulators. \href{http://dx.doi.org/10.1038/nphoton.2014.9}{{Nature Photonics}, 8(2014), 229--233.}
		
	\bibitem{Haffner2016}
	Haffner, C., Heni, W., Fedoryshyn, Y., Josten, A., Baeuerle, B., Hoessbacher, C., Salamin, Y., Koch, U., Dordevic, N., Mousel, P., Bonjour, R., Emboras, A., Hillerkuss, D., Leuchtmann, P., Elder, D.~L., Dalton, L.~R., Hafner, C., Leuthold, J. 2016. Plasmonic organic hybrid modulators---scaling highest speed photonics to the microscale. \href{http://dx.doi.org/10.1109/JPROC.2016.2547990}{{Proceedings of the IEEE}, 104(2016), 2362--2379.}
	
	
	\bibitem{Hoessbacher2017}
	Hoessbacher, C., Josten, A., Baeuerle, B., Fedoryshyn, Y., Hettrich, H., Salamin, Y., Heni, W., Haffner, C., Kaiser, C., Schmid, R., Elder, D.~L., Hillerkuss, D., M\"{o}ller, M., Dalton, L.~R., Leuthold, J. 2017. Plasmonic modulator with >170 GHz bandwidth demonstrated at 100 GBd NRZ. \href{http://www.opticsexpress.org/abstract.cfm?URI=oe-25-3-1762}{{Optics Express}, 25(2017), 1762--1768.}
	
		
	\bibitem{Tian2009}
	Tian, J., Yu, S., Yan, W., Qiu, M. 2009. Broadband high-efficiency
	surface-plasmon-polariton coupler with silicon-metal interface.
	\href{http://dx.doi.org/10.1063/1.3168653}{{Applied Physics Letters}, 95(2009), 013504.}
	
	
	\bibitem{Han2010}
	Han, Z., Elezzabi, A.~Y., Van, V. 2010. Experimental realization of subwavelength plasmonic slot waveguides on a silicon platform. \href{http://ol.osa.org/abstract.cfm?URI=ol-35-4-502}{{Optics Letters}, 35(2010), 502--504.}
	
	
	\bibitem{Chen2015}
	Chen, C.-T., Xu, X., Hosseini, A., Pan, Z., Subbaraman, H., Zhang, X., Chen, R. T. 2015. Design of highly efficient hybrid Si-Au taper for dielectric strip waveguide to plasmonic slot waveguide mode converter. \href{http://dx.doi.org/10.1109/JLT.2015.2390040}{{Journal of Lightwave Technology}, 33(2015), 535--540.}
	
	
	\bibitem{Zhu2016a}
	Zhu, B.~Q., Tsang, H.~K. 2016. High coupling efficiency silicon waveguide to
	metal--insulator--metal waveguide mode converter. \href{http://jlt.osa.org/abstract.cfm?URI=jlt-34-10-2467}{{Journal of Lightwave 	Technology}, 34(2016), 2467--2472.}
	
	
	\bibitem{Ono2016a}
	Ono, M., Taniyama, H., Xu, H., Tsunekawa, M., Kuramochi, E., Nozaki, K.,	Notomi, M. 2016. Deep-subwavelength plasmonic mode converter with large size reduction for Si-wire waveguide. \href{http://www.osapublishing.org/optica/abstract.cfm?URI=optica-3-9-999}{{Optica}, 3(2016), 999--1005.}
	
	
	\bibitem{Veronis2007a}
	Veronis, G., Fan, S.~H. 2007. Modes of subwavelength plasmonic slot waveguides. \href{http://dx.doi.org/10.1109/JLT.2007.903544}{{Journal of Lightwave Technology}, 25(2007), 2511--2521.}
	
	
	\bibitem{Kewes2016}
	Kewes, G., Schoengen, M., Neitzke, O., Lombardi, P., Schönfeld, R.-S.,	Mazzamuto, G., Schell, A.~W., Probst, J., Wolters, J., Löchel, B., Toninelli, C., Benson, O. 2016. A realistic fabrication and design concept for quantum gates based on single emitters integrated in plasmonic-dielectric waveguide structures. \href{http://dx.doi.org/10.1038/srep28877}{{Scientific Reports}, 6(2016), 28877.}
	
	
	\bibitem{Palik1985}
	Palik, E.~D. 1985. {Handbook of optical constants of solids}. Vol.~1. Academic Press, London, 804s. 
	
	\bibitem{Kitamura2007}
	Kitamura, R., Pilon, L., Jonasz, M. 2007. Optical constants of silica glass from extreme ultraviolet to far infrared at near room temperature. \href{http://ao.osa.org/abstract.cfm?URI=ao-46-33-8118} {{Applied Optics}, 46(2007), 8118--8133.}
		
	\bibitem{McPeak2015}
	McPeak, K.~M., Jayanti, S.~V., Kress, S.~J.~P., Meyer, S., Iotti, S., Rossinelli, A., Norris, D.~J. 2015. Plasmonic films can easily be better: Rules and recipes. \href{http://dx.doi.org/10.1021/ph5004237}{{ACS Photonics}, 2(2015), 326--333.}	
	
	\bibitem{Yang2016a}
	Yang, J., Hugonin, J.-P., Lalanne, P. 2016. Near-to-far field transformations for	radiative and guided waves. \href{http://dx.doi.org/10.1021/acsphotonics.5b00559}{{ACS Photonics}, 3(2016), 395--402.}	
	
	\bibitem{Ji2017}
	Ji, X., Barbosa, F.~A.~S., Roberts, S.~P., Dutt, A., Cardenas, J., Okawachi, Y., Bryant, A., Gaeta, A.~L., Lipson, M. 2017. Ultra-low-loss on-chip resonators with sub-milliwatt parametric oscillation threshold. \href{http://www.osapublishing.org/optica/abstract.cfm?URI=optica-4-6-619}{{Optica}, 4(2017), 619--624.}	
	
	\bibitem{Veronis2007}
	Veronis G., Fan, S. 2007. Theoretical investigation of compact couplers between	dielectric slab waveguides and two-dimensional metal-dielectric-metal plasmonic waveguides. \href{http://www.opticsexpress.org/abstract.cfm?URI=oe-15-3-1211}{{Optics Express}, 15(2007), 1211--1221.} 

	
\end{thebibliography}

\end{document}